\documentclass[10pt,aps,prd,nofootinbib,showpacs,eqsecnum,superscriptaddress,longbibliography,preprintnumbers]{revtex4-2}
\usepackage{graphicx}
\usepackage{amsmath,amssymb}
\usepackage{amsfonts}
\usepackage{xspace} 
\usepackage[usenames]{color}
\usepackage{dcolumn}
\usepackage{bm}
\usepackage{mathrsfs}
\usepackage[utf8]{inputenc} 
\usepackage{multirow}
\usepackage{etoolbox}
\usepackage{tikz}
\usepackage[normalem]{ulem}
\usepackage{cancel}
\usepackage[colorlinks=true]{hyperref}
\usepackage[all]{hypcap}

\newcommand{\dd}{\mathrm{d}}

\begin{document}

\title{Rotating black holes with primary hair in five-dimensional generalized Proca theory}

\author{Mokhtar Hassaine}
\email{hassaine@inst-mat.utalca.cl}
\affiliation{Instituto de Matemáticas, Universidad de Talca, Casilla 747, Talca, Chile}

\author{Ulises Hernandez-Vera}
\email{uhernandez.vera@gmail.com}
\affiliation{Instituto de Matemáticas, Universidad de Talca, Casilla 747, Talca, Chile}

\begin{abstract}

This work presents a new class of exact analytic rotating black hole solutions within five-dimensional generalized Proca theories. Through a Kerr-Schild ansatz where the Proca field is set along a null geodesic congruence, the non-linear field equations reduce to a consistent set of three master equations. This geometric configuration ensures that the vector field remains light-like on-shell, effectively restricting the theory’s functional couplings to discrete constants and allowing for a fully analytic treatment. The resulting solutions, incorporating a cosmological constant and two independent angular momenta, exhibit primary hair given by an arbitrary function of the non-Killing angular coordinate. We identify several solution branches defined by specific algebraic relations between the Proca coupling constants, providing a significant generalization of the Myers-Perry family. Notably, the metric retains a Kerr-Schild form identical to the Myers-Perry representation, with an additional contribution constructed from the tensor product of the Proca one-form with itself.

\end{abstract}

\maketitle

\section{Introduction}

The inception of black hole physics can be traced back to the discovery of the Schwarzschild solution, a foundational milestone that revealed the static structure of spacetime around a spherical mass. However, the path toward understanding its rotating counterpart proved to be an exceptionally arduous one, spanning nearly half a century before the Kerr metric was finally unveiled \cite{Kerr:1963ud}. This prolonged gap in the history of General Relativity underscores the formidable nature of the Einstein field equations which, even under the simplifying assumptions of stationarity and axial symmetry, remain a highly non-linear system of coupled partial differential equations. A transformative breakthrough in this field occurred with the realization that the Kerr metric, and subsequently the charged Kerr-Newman solution \cite{Newman:1965my}, could be expressed through the elegant and structurally powerful Kerr-Schild formulation, see \cite{Kerr:1965vyg} and \cite{Debney:1969zz}. This framework expresses the physical metric as a linear perturbation of a background or "seed" metric through the introduction of a scalar profile function and a null, geodesic vector field. In these classical cases, the seed metric is Minkowski space, and the geometry is generated by a null, geodesic, and shear-free congruence. According to the Kerr theorem, an infinite variety of analytic functions can produce these congruences, meaning that selecting the precise function for the Kerr geometry typically requires knowing the solution beforehand. A stationary-axisymmetric version of the theorem formulated in \cite{Ayon-Beato:2015nvz} resolves this degeneracy. By establishing a rigorous, symmetry-based selection rule, this framework systematically identifies the unique congruence underpinning the Kerr solution, eliminating the need for ad hoc assumptions.

The utility of the linear Kerr-Schild structure extends well beyond four dimensions. Indeed, the Myers-Perry solution \cite{Myers:1986un} and its generalizations with a cosmological constant \cite{Gibbons:2004js, Gibbons:2004uw}, which extend the Kerr-(A)dS geometry to higher dimensions with multiple rotation planes, also admit a Kerr-Schild representation. This indicates that the ansatz captures a fundamental property of the gravitational field for rotating objects, effectively linearizing the field equations regardless of the spacetime dimensionality. While the Kerr-Schild ansatz extends straightforwardly to higher-dimensional vacuum solutions \cite{Myers:1986un} or those with a cosmological constant \cite{Gibbons:2004js, Gibbons:2004uw}, the inclusion of matter fields represents a completely different challenge. A major hurdle in this direction is the derivation of a five-dimensional analogue of the Kerr-Newman solution, which remains one of the most persistent open problems in the field. Within pure Einstein-Maxwell theory, no general exact analytic solution for a rotating charged black hole has been found to date. This gap in our knowledge is only partially bridged by the realization that the inclusion of a Chern-Simons term (a hallmark of five-dimensional supergravity) allows for the existence of charged rotating solutions \cite{Chong:2005hr}. It is worth noting that, in many instances, these solutions were originally derived through alternative methods rather than the Kerr-Schild ansatz itself. However, it was subsequently demonstrated that they can indeed be cast into a generalized or extended Kerr-Schild framework, which necessitates the introduction of two scalar profile functions and a spacelike vector field orthogonal to the null geodesic field \cite{Aliev:2008bh}. A pivotal observation in this direction was recently highlighted by the work in Ref. \cite{Fernandes:2026rjs}, which recognized that the efficacy of the Kerr-Schild ansatz can be preserved even when navigating complex matter content. Specifically, by enforcing a configuration where the Proca field is strictly aligned with the null geodesic congruence of the geometry, Fernandes demonstrated that the ansatz remains remarkably powerful for deriving exact analytic solutions within a broad class of generalized Proca theories \cite{Fernandes:2026rjs}. These solutions are of particular interest because they possess "primary hair" and break the circularity condition. For related developments in four dimensions, we also refer to recent works on black holes with primary scalar hair \cite{Bakopoulos:2023fmv, Baake:2023zsq}. Note that the existence of the hairy black hole solutions presented here does not contradict the classical no-hair theorems. These theorems are established under assumptions that are specific to four-dimensional spacetimes, and therefore do not directly apply in the present five-dimensional framework.

The core objective of this work is to determine whether the framework developed in \cite{Fernandes:2026rjs} for four-dimensional generalized Proca theories can be extended to higher dimensions, taking the five-dimensional case as a first step. This extension is far from automatic; the transition from four to five dimensions introduces significant structural challenges. A classic example of this sensitivity is found in Einstein-Maxwell theory, where the Kerr-Schild ansatz successfully yields the $4$D Kerr-Newman solution but fails to produce its rotating, charged counterpart in $5$D. This demonstrates how strongly the analytic tractability of these systems depends on both dimensionality and gauge field couplings. Furthermore, five dimensions introduce a more intricate rotational structure due to the existence of two independent angular momenta. To navigate this technical complexity, we leverage known higher-dimensional Kerr-Schild representations of vacuum or (A)dS solutions. Our goal is to test whether the robust features of the $4$D solutions, namely their analytic tractability and the emergence of primary hair, persist in this 5D environment.

The remainder of this paper is organized as follows. In Sec.~II, we define the five-dimensional generalized Proca action. Sec.~II-A is devoted to the implementation of the Kerr-Schild ansatz, where we introduce the five-dimensional seed metric and the null geodesic congruence. We derive the master field equations and highlight the resulting nested Kerr-Schild structure. The analytic solutions are presented across different coupling regimes in Sec.~II-B, followed by an analysis of the event horizon properties and the role of primary hair in Sec.~II-C. We also extend our framework to the toroidal ($\kappa=0$) case and provide a formal proof of the mathematical obstruction that arises when the standard Maxwell kinetic term (or its non linear extension) is included in five dimensions. Sec.~III is devoted to our conclusions and perspective. 

\section{Five-dimensional generalized Proca theory and its Kerr-Schild ansatz}

In this work, we explore the gravitational dynamics of a five-dimensional vector field $A_\mu$ non-minimally coupled to curvature within the framework of generalized Proca theories. The action under consideration is generically defined  by 
\begin{equation}
    S = \int d^{5}x \sqrt{-g} \left( \sum_{n=2}^{4} \mathcal{L}_{n} \right), \label{procaLgeneral}
\end{equation}
where the Lagrangian is decomposed into three distinct sectors, $\mathcal{L}_2, \mathcal{L}_3,$ and $\mathcal{L}_4$, each representing different orders of derivatives and non-minimal couplings. The specific functional forms of the Lagrangians read
\begin{equation}\label{eq:Lall}
\begin{aligned}
\mathcal{L}_2 &= G_2(X),\\[4pt]
\mathcal{L}_3 &= G_3(X)\,\nabla_\mu A^\mu,\\[4pt]
\mathcal{L}_4 &= G_4(X)\,R + G_{4 X}(X)\!\left[\,(\nabla_\mu A^\mu)^2 - \nabla_\rho A_\sigma\,\nabla^\sigma A^\rho\right],\\[4pt]
\end{aligned}
\end{equation}
where $X \equiv -A_{\mu}A^{\mu}/2$ denotes the norm of the vector field. The functions $G_n(X)$ with $n=2, 3$ and $4$ are arbitrary functions of this norm, and $G_{n X}$ represents the derivative of $G_n$ with respect to $X$. It is important to emphasize that this action represents a specific and physically motivated sub-class of the much broader family of theories described in the seminal work on the generalization of the Proca action  \cite{Heisenberg:2014rta}. By construction, this structure ensures that the resulting equations of motion remain strictly second-order for both the metric and the vector field. This property is crucial as it avoids the introduction of ghost-like degrees of freedom associated with Ostrogradski instabilities, which typically plague higher-derivative theories, while simultaneously maintaining a non-trivial non-minimal coupling to the Ricci scalar $R$.

Our investigation of these specific Lagrangians is strongly motivated by earlier works \cite{Babichev:2017rti} as well as by recent developments in the study of regularized Proca theories in four dimensions. In the four-dimensional case, it has been demonstrated that such theories admit static and spherically symmetric black hole solutions characterized by the presence of primary hair \cite{Charmousis:2025jpx}. Note that the structure of these Proca Lagrangians is conceptually similar to the regularized frameworks of four-dimensional Gauss-Bonnet gravity \cite{Glavan:2019inb, Fernandes:2020nbq, Hennigar:2020lsl, Fernandes:2021dsb, Fernandes:2022zrq}. Another surprising and notable feature of these generalized Proca theories is the existence of static black hole solutions for which a cosmological constant appears as a constant of integration. This provides an effective vacuum energy that governs the asymptotic behaviour of the spacetime even in the complete absence of a bare cosmological term in the original action, suggesting a novel mechanism for self-acceleration or de Sitter-like expansions \cite{Charmousis:2026dbi}. Interest in these generalized Proca theories can be further explored in several recent works. Specifically, the construction of exact rotating black holes with primary hair was recently advanced in \cite{Fernandes:2026rjs}. Additionally, the observational and perturbative signatures of such Proca hair, including double-peak optics and echoes, have been analysed in \cite{Konoplya:2025uiq, Konoplya:2025bte}. Further extensions involving regularized Lovelock-Proca corrections and their cosmological implications can be found in \cite{Fernandes:2025mic}. To conclude, let us mention that there also exists a more general action than \eqref{procaLgeneral} involving, for example, a possible coupling to the Gauss-Bonnet density $\mathcal{G}$. In this context, a recent study \cite{Kleihaus:2026rev} has explored spontaneously vectorized black holes in Einstein-vector-Gauss-Bonnet theory, demonstrating that these solutions can carry electric charges or magnetic dipole moments and typically possess lower free energy than their general relativistic counterparts.

It is imperative to highlight a deliberate and significant choice in our current formulation, namely the action presented in Eq.~(\ref{procaLgeneral}) does not include the standard Maxwell kinetic term, $F_{\mu\nu}F^{\mu\nu}$. Its absence here is not an oversight but a strategic necessity for the analytic framework we are developing. In the context of the Kerr-Schild ansatz and the specific vector configurations we seek to extend to five dimensions, the inclusion of the standard Maxwell field strength introduces a severe mathematical obstruction. This obstruction manifests as a fundamental incompatibility that arises specifically in higher dimensions when one insists on a vector field configuration where the potential is proportional to the null geodesic congruence ($A \propto l$). Unlike the four-dimensional case, where such an alignment is perfectly consistent with the Maxwell equations and allows for the derivation of the Kerr-Newman solution, this is no longer true in five dimensions. We shall provide a detailed discussion of this obstruction in the subsequent sections, demonstrating why the ``Proca-style'' derivative couplings used here allow for a degree of integrability and analytic tractability that is lost when the standard Maxwell kinetic term is strictly enforced. By avoiding the Maxwell term while keeping the theory ghost-free through the generalized Proca framework, we open a window into a new class of rotating solutions in five dimensions that would otherwise be probably inaccessible.

The field equations obtained by varying the action with respect to the metric $g^{\mu\nu}$ and the vector field $A_\mu$ respectively read $ E_{\mu\nu}=0$ and $\mathcal{J}_A^{\mu}=0$ where 
\begin{equation}
\begin{aligned}
        E_{\mu\nu}:&=\mathcal{E}^{(G_2)}_{\mu\nu}+\mathcal{E}^{(G_3)}_{\mu\nu}+\mathcal{E}^{(G_4)}_{\mu\nu}, \\
        \mathcal{E}^{(G_2)}_{\mu\nu}:&=  -\frac{1}{2}\bigg(g_{\mu \nu} \,G_2\,+\,A_{\mu}\,A_{\nu}\,G_{2X} \bigg),\\
        \mathcal{E}^{(G_3)}_{\mu\nu}:&=  \frac{G_{3X}}{2}\bigg(2\nabla_{(\mu}\left[A^{\alpha}\right]\,A_{\nu)}\,A_{\alpha}-A_{\mu}A_{\nu}\nabla_{\alpha}A^{\alpha}-g_{\mu\nu}\,A^{\alpha}A^{\beta}\,\nabla_{\beta}A_{\alpha}\bigg),
        \\
        \mathcal{E}^{(G_4)}_{\mu\nu}:&=G_4\,G_{\mu\nu}+G_{4X}\Bigg(R_{(\mu |\alpha|}A_{\nu)}A^{\alpha}-g_{\mu\nu}R_{\alpha\beta}\,A^{\alpha}A^{\beta}+R_{\mu \alpha\nu \beta}A^{\alpha}A^{\beta}-\frac{1}{2}A_{\mu}\,A_{\nu}\,R\\
        &\qquad +\frac{g_{\mu \nu}}{2} \nabla_{\alpha}\nabla_{\beta}\left(A^{\alpha}A^{\beta}\right)- \nabla_{(\mu}\nabla_{\beta}\left[A_{\nu)}\,A^{\beta}\right]+\frac{\square\left(A_{\mu}\,A_{\nu}\right)}{2}+g_{\mu\nu}\square\left(X\right)-\nabla_{\mu}\nabla_{\nu}\left(X\right)\Bigg)\\
        &\quad+G_{4XX}\Bigg(\frac{1}{2}A_{\mu}\,A_{\nu}\left[\left(\nabla_{\alpha}A_{\beta}\right)^2-\left(\nabla\,A\right)^2\right]-\left[A_{(\mu}\,\nabla^{\beta}\,A_{\nu)}+A_{(\mu}\nabla_{\nu)}A^{\beta}\right]\,A^{\alpha}\nabla_{\alpha}A_{\beta}\\
        &\qquad+ 2\,g_{\mu \nu} \nabla_{\beta}(A^{[\beta})\,A^{\alpha]}\nabla_{\alpha}(X)-2\,\nabla_{(\mu}(X)\,A_{\nu)} \nabla_{\beta}(A^{\beta})-\nabla_{(\mu}\,A_{\nu)} \,\nabla_{\alpha}(X)\,A^{\alpha}-\nabla_{\mu}(X)\nabla_{\nu}(X)\Bigg),
        \\
        \mathcal{J}_A^{\mu}:=&-A^{\mu}G_{2X}+(A^{\alpha}\nabla^{\mu}(A_{\alpha})-\nabla^{\alpha}(A_{\alpha})\,A^{\mu})G_{3X}+2G^{\mu}_{\,\,\alpha}A^{\alpha}G_{4X}\\
        &+\Bigg(A^{\mu}(\nabla_{\alpha}A_{\gamma})^2-A^{\mu}(\nabla_{\alpha}A^{\alpha})^2+2A^{
    \alpha}[\nabla_{\sigma}A^{\sigma}\nabla^{\mu}A_{\alpha}-\nabla_{\sigma}A_{\alpha}\nabla^{\mu}A^{\sigma}]G_{4XX}\Bigg).
\end{aligned}
\end{equation}

\subsection{The Kerr-Schild ansatz}

To address the analytical challenges posed by the field equations in five-dimensional generalized Proca theories, we employ a solution strategy based on the Kerr-Schild framework.  Our construction assumes that the physical spacetime geometry can be decomposed into a background or seed metric, modified by a linear perturbation that depends on a scalar profile function and a specific vector field. In doing so, we define the  following  seed metric 
\begin{eqnarray}
\label{seedm}
d {s}_0^2= -\frac{\left(1-\lambda r^2\right) {\Delta}_{\theta}}{\Xi_a \Xi_b} d t^2+\frac{{\Sigma}(r, \theta) r^2}{\left(1-\lambda r^2\right)\left(r^2+ a^2\right)\left(r^2+ b^2\right)} d r^2+\frac{{\Sigma}(r,\theta)}{{\Delta}_{\theta}} d \theta^2+\frac{\left(r^2+a^2\right) \sin^2(\theta) }{{\Xi}_a} d \phi^2+\frac{\left(r^2+ b^2\right) \cos ^2( \theta)}{{\Xi}_b} d \psi^2,
\end{eqnarray}
and, where for simplicity, we have defined 
\begin{equation}
{\Delta}_\theta= 1 +\lambda a^2\cos^2(\theta) +\lambda b^2\sin^2(\theta),\qquad
{\Sigma}(r,\theta) = r^2 + a^2\cos^2(\theta) + b^2\sin^2(\theta), \quad {\Xi}_a = 1 +\lambda a^2,\quad {\Xi}_b = 1 +\lambda b^2.
\label{metricfun}
\end{equation}
It is straightforward to verify that the seed metric \eqref{seedm} satisfies the five-dimensional Einstein field equations in the presence of a cosmological constant $\lambda$, namely
\begin{equation}
    G_{\mu\nu} + 6\lambda g_{\mu\nu} = 0.
\end{equation}
In order to implement the Kerr-Schild ansatz, we also need a null and geodesic vector which reads in this case
\begin{eqnarray}
    \label{nullvect2}
    \ell= \frac{\Delta_\theta}{\Xi_a \Xi_b}\ dt + \frac{r^2 \Sigma(r,\theta)}{(1-\lambda r^2)(r^2+ a^2)(r^2+ b^2)}\, dr  - \frac{a}{\Xi_a} \sin^2(\theta) d\phi - \frac{b}{\Xi_b}\cos^2( \theta) d\psi.\nonumber
\end{eqnarray}

Leveraging the geometric properties of the seed metric and the associated null geodesic congruence, we adopt the following extended Kerr-Schild ansatz inspired by the four-dimensional case \cite{Fernandes:2026rjs}
\begin{eqnarray}
\label{KSansatz}
  d s^2=d s_0^2+h_1(r,\theta)\, \ell \otimes \ell,\qquad  A=h_2(r,\theta)\,\ell. 
\end{eqnarray}
By implementing this specific ansatz, where the Proca field is strictly aligned with the null generators of the spacetime, the complexity of the field equations is dramatically reduced. A fundamental consequence of this alignment is that the vector field norm, $X \equiv -A_{\mu}A^{\mu}/2$, vanishes identically on-shell. Note that the scalar \(X\) also vanishes for the four-dimensional Kerr--Newman solution, as well as for the known five-dimensional charged rotating black hole solution of Einstein--Maxwell--Chern--Simons theory \cite{Aliev:2008bh}. Because all the coupling functions $G_n(X)$ and their derivatives are evaluated exactly at $X=0$, these expressions collapse from arbitrary functional forms into discrete constants, rather than maintaining any non-trivial dependence on the coordinates $r$ and $\theta$. Moreover, as in the four-dimensional case \cite{Fernandes:2026rjs} the complete set of coupled Einstein-Proca equations can be condensed into a linear combination of just three fundamental master equations, two of them being exact total derivatives
\begin{align}
E_{rr} &:=\frac{\partial}{\partial r}\!\bigg\{
\Sigma\,
\Bigl(
h_2^2\,G_{4X}-G_{4}\,h_1
\Bigr)-P(r,\theta)\left( 12\,G_{4}\,\lambda + G_2 \right)
\bigg\}\\[6pt]
E_{\theta\theta} &:=\frac{\partial}{\partial \theta}\!\bigg\{
\Sigma\,
\Bigl(
h_2^2\,G_{4X}-G_{4}\,h_1
\Bigr)
\bigg\} , \\[6pt]
\mathcal{J}^{r} &= \left[
\frac{\partial}{\partial r}\Big(\Sigma\,G_{4X}\,h_1\Big)
+\frac{\Sigma}{\Sigma+2r^2}
\left(
\frac{\partial}{\partial r}\Big(\Sigma\,r\,h_2\,G_{3X}\Big)+\frac{\partial}{\partial r}\Big((\Sigma+2r^2)\,h_2^2\,G_{4XX}\Big)
+r\,\Sigma\,(G_{2X}+\lambda\,G_{4X})
\right)
\right],
\label{Jrr}
\end{align}
where 
\begin{eqnarray}
\label{prt}
    P(r,\theta)=\int \frac{\Sigma(r,\theta)^2\,r}{\Sigma(r,\theta)+2\,r^2}\dd r. 
\end{eqnarray}
Consequently, this triad of equations captures the entire dynamical content of the theory under the chosen ansatz. Ensuring that these specific relations are satisfied is sufficient to guarantee that the full system of field equations is identically fulfilled, thereby reducing a formidable set of non-linear partial differential equations to a more manageable integrable core.

Now, since $\partial^2_{r\theta}P(r,\theta)\not=0$, the two first equations $E_{rr}=0=E_{\theta\theta}$ will be compatible if and only if 
\begin{equation}
G_2(X) = -12 \lambda G_{4}(X),\label{relationg2yg4}
\end{equation}
and, hence these first two equations will be satisfied provided that   
\begin{equation}
h_1(r,\theta)=\frac{M}{\Sigma}+\frac{h_2(r,\theta)^2\,G_{4X}}{G_{4}},\label{h_1eq}
\end{equation}
with $M$ an integration constant proportional to the mass. It is particularly illuminating to observe that the specific form of the metric function $h_1$, as given by \eqref{h_1eq}, allows the geometry to be expressed through a nested Kerr-Schild decomposition. Specifically, our ansatz can be recast in the following suggestive form
\begin{equation}
\label{KSansatz2}
d s^2 = d s_0^2 + \frac{M}{\Sigma}\, \ell \otimes \ell + \frac{G_{4X}}{G_4} A \otimes A, \qquad A = h_2(r, \theta) \, \ell.
\end{equation}
This representation highlights a remarkable structural hierarchy. The first part of the metric, $d s_{MP}^2 = d s_0^2 + \frac{M}{\Sigma}\, \ell \otimes \ell$, is precisely the Kerr-Schild representation of the (AdS)-Myers-Perry solution \cite{Myers:1986un, Gibbons:2004js, Gibbons:2004uw}. In this light, our solution can be interpreted as a further linear deformation of the Myers-Perry background, where the "perturbing" agent is the Proca one-form $A$. This nesting is possible because the Proca field is strictly aligned with the same null congruence $\ell$ that generates the rotating vacuum geometry.

The remaining independent field equation, $\mathcal{J}^{r}=0$, serves to uniquely determine the functional form of the profile $h_2(r, \theta)$, equation that reads
\begin{equation}
    \mathcal{J}^r =  \left[ \frac{\partial}{\partial r}\left(\frac{\Sigma\,G_{4X}^2 h^2_2}{G_4}\right) + \frac{\Sigma}{\Sigma+2r^2} \left( \frac{\partial}{\partial r}\left(\Sigma r h_2 G_{3X}\right) + \frac{\partial}{\partial r}\left((\Sigma+2r^2) h_2^2 G_{4XX}\right) \right) \right]. \label{corrientesinG2=G4}
\end{equation}
To solve this equation analytically, we investigate the behaviour of the generalized Proca action in the vicinity of the vanishing kinetic norm, $X=0$. Under the assumption that the coupling functions $G_3(X)$ and $G_4(X)$ are sufficiently smooth to admit a Taylor series expansion, and identifying $G_2$ through the compatibility constraint \eqref{relationg2yg4}, we consider the following expansions
\begin{eqnarray}
\label{Gn_expansions_final}
    G_3(X) = \alpha c_{31} X + \mathcal{O}(X^2),\quad G_4(X) = 1 + \alpha c_{41} X + \alpha^2 c_{42} X^2 + \mathcal{O}(X^3),\quad  G_2(X) = -12\lambda G_4(X).
\end{eqnarray}
Crucially, we contend that higher-order powers of $X$ beyond those explicitly presented in \eqref{Gn_expansions_final} do not contribute to the essential physics of the model. By substituting the perturbative expansions for the coupling functions $G_n(X)$ into \eqref{corrientesinG2=G4}, the governing equation for the profile function $h_2(r, \theta)$ can be recast as a first-order non-linear  differential equation
\begin{equation}
\label{remaineq}
   \frac{\partial h_2}{\partial r} + \frac{ \alpha h_2 c_{31} \Sigma (\Sigma + 2r^2) + 6r h_2^2 \alpha^2 \left[ 2c_{42} \Sigma + \frac{1}{3} c_{41}^2 (\Sigma + 2r^2) \right] }{ \Sigma \left[ 2h_2 \alpha^2 (\Sigma + 2r^2) (2c_{42} + c_{41}^2) + \Sigma r \alpha c_{31} \right] } = 0.
\end{equation}

\subsection{Analytic solutions for different coupling regimes }\label{subsec:Analytic}

To obtain reasonably simple analytic forms for the remaining field equation \eqref{remaineq}, we restrict our analysis to specific parameter space configurations. By imposing particular algebraic relations among the coupling constants $\{c_{31}, c_{41}, c_{42}\}$, we are able to reduce the governing first-order differential equation to integrable cases. This approach follows the logic successfully applied in the four-dimensional context \cite{Fernandes:2026rjs, Cisterna:2016nwq}, where such constraints allowed for the discovery of closed-form solutions that clearly illustrate the physical impact of the Proca hair. In the following, we explore three relevant regimes that lead to distinct classes of analytic expressions. Notably, across all three cases, the system consistently preserves an arbitrary function denoted $\mathcal{F}_1(\theta)$. This functional freedom arises because the governing partial differential equation \eqref{remaineq} involves exclusive radial derivative, leaving the angular dependence unfixed. Crucially, as shown below, this remaining freedom will prove instrumental in ensuring the existence of a well-defined event horizon.

\begin{itemize}
   \item \textbf{Case I: The coupling choice \texorpdfstring{$c_{42} = -c_{41}^2/2$}}. \\
This specific parameter choice simplifies the structural complexity of the equation \eqref{remaineq}. Indeed, enforcing the relation $c_{42} = -c_{41}^2/2$ cancels a dominant term in the denominator of the governing differential equation, significantly reducing its algebraic structure. Upon integration, this simplified differential equation yields a solution characterized by logarithmic dependences, 
    \begin{equation}
        h_2 (r,\theta)= \frac{c_{31} \Sigma (r^2 - \Sigma)^2}{r \left[ 2\alpha c_{41}^2 \Sigma^2 \ln(\Sigma) + \mathcal{F}_1(\theta) c_{31} \Sigma^2 (r^2 - \Sigma)^2 - 4\alpha c_{41}^2 \left( \Sigma^2 \ln(r) + \frac{3}{4}(r^2 - \Sigma)(\frac{r^2}{3} - \Sigma) \right) \right]}.\label{h2casoI}
    \end{equation}
   
  \item \textbf{Case II: Vanishing linear $G_4$ derivative \texorpdfstring{$c_{41} = 0$}}. {\textbf  {Stealth on (AdS)-Myers-Perry black hole.}} \\
    When the linear correction to the $G_4$ function is suppressed \eqref{Gn_expansions_final}, from equation \eqref{h_1eq}, we have 
    \[
    h_1(r,\theta)=\frac{M}{\Sigma},
    \]
    and this is nothing but the Kerr-Schild function for the (AdS)-Myers-Perry black hole solution in five dimensions. Hence, the solution reduces to a stealth defined in the Myers-Perry spacetime where the vector potential $A=h_2(r,\theta)\,\ell$ with  
    \begin{equation}
        h_2(r, \theta) = \frac{ \sqrt{c_{31}^2 r^2 \Sigma^2 - 8\alpha c_{42} \mathcal{F}_1(\theta) (\Sigma + 2r^2)} - c_{31} r \Sigma }{ 4 \alpha c_{42} (\Sigma + 2r^2) }.
    \end{equation}
As in the four-dimensional case, this solution turns out to be circular and generalizes those found in four dimensions in \cite{Fernandes:2026rjs} and \cite{Cisterna:2016nwq}.  

    \item \textbf{Case III: Vanishing $G_3$ sector, \texorpdfstring{$c_{31} = 0$}.} \\
    In the absence of the $G_3$ coupling constant, the field equation simplifies to a separable form, yielding a power-law solution
    \begin{equation}
    \label{dfg}
        h_2(r, \theta) = \mathcal{F}_1(\theta) \, \Sigma^{-\left(\frac{c_{41}^2}{2(2c_{42}+c_{41}^2)}\right)} \left[ \Sigma + 2r^2 \right]^{\left(-\frac{c_{42}}{2c_{42}+c_{41}^2}\right)}.
    \end{equation}
    This solution highlights how the Proca hair scales with the radial structural function $\Sigma$ under the influence of higher-order $G_4$ corrections. 
    \begin{itemize}
        \item \textbf{Case III-a: Vanishing $G_3$ sector, \texorpdfstring{$c_{31} = 0$ with $c_{42}=0$}.} \\
       It is particularly interesting to analyse the specific case where $c_{42}=0$ in addition to $c_{31} = 0$, which leads to the simplification $h_2 = \mathcal{F}_1(\theta)/\sqrt{\Sigma}$. Under these conditions, the nested Kerr-Schild structure \eqref{KSansatz2} collapses into a single-vector deformation of the seed metric
\begin{equation}
\label{KSansatz22}
d s^2 = d s_0^2 + \frac{M + \alpha c_{41}\mathcal{F}_1(\theta)^2}{\Sigma}\, \ell \otimes \ell, \qquad A = \frac{\mathcal{F}_1(\theta)}{\sqrt{\Sigma}} \, \ell.
\end{equation}
This simplified form reveals two distinct physical regimes depending on the nature of the arbitrary function $\mathcal{F}_1(\theta)$. First, if we consider $\mathcal{F}_1(\theta) = q$ (a constant), the metric reduces exactly to the  (AdS)-Myers-Perry solution, but with a shifted mass parameter $M_{eff} = M + \alpha c_{41} q^2$. In this regime, the Proca field behaves as a "stealth" configuration; it possesses its own dynamics and supports the geometry, yet the resulting spacetime remains indistinguishable from the standard vacuum solution. Second, when $\mathcal{F}_1(\theta)$ is a non-trivial function of the angular coordinate, the solution departs from the vacuum (or AdS) Myers-Perry family. While the metric maintains the formal structure of a standard Kerr-Schild ansatz, the $\theta$-dependence in the scalar function introduces a profound shift in the underlying symmetries. Specifically, this non-constant primary hair breaks the circularity condition, providing a concrete example of how the generalized Proca terms allow for analytic, rotating configurations that bypass the rigid symmetry constraints typically imposed on stationary black holes in General Relativity.

\end{itemize}

\end{itemize}

\subsection{Some properties of the solutions }

In this subsection, we begin by investigating the conditions required for the existence of the horizon and analyze the constraints that arise from the field equations. Establishing the existence of a well-defined horizon is a crucial step, as it determines whether the solutions under consideration can be interpreted as physically relevant black hole configurations. We therefore first examine the general structure of the solutions and identify the requirements that allow the horizon to emerge consistently within our framework.

Once these conditions have been established, we specialize our analysis to the case of a toroidal horizon geometry. This particular topology is of special interest because it can exhibit properties that differ significantly from those associated with more conventional spherical configurations. We study how the horizon structure and the corresponding field configurations are modified in this setting and discuss the implications of adopting this topology for the resulting solutions.

Finally, we investigate the consequences of introducing an additional Maxwell term into the theory. As will be shown, the inclusion of this contribution substantially modifies the structure of the equations and disrupts the mechanism that allowed the previous construction to exist. Our analysis indicates that the conditions required for obtaining consistent solutions can no longer be simultaneously satisfied. This suggests that solutions of the type discussed in the previous cases cease to exist once the Maxwell sector is incorporated, at least within the particular setup and assumptions adopted throughout this work.

\subsubsection{Existence of an horizon}

Regarding the existence and location of the event horizon for these solutions, we restrict ourselves, for simplicity, to the asymptotically flat case ($\lambda=0$). In the absence of circularity, the location of the event horizon cannot, in general, be characterized by a surface of constant radial coordinate. Instead, the horizon must be described by a more general hypersurface of the form\footnote{We are grateful to Pedro Fernandes for drawing our attention to this subtle point.}
\[
F(r,\theta):=r-r_H(\theta)=0,
\]
where the function $r_H(\theta)$ determines the angular dependence of the horizon position \cite{Babichev:2025szb}. The horizon condition is then obtained by requiring this hypersurface to be null, namely
\[
g^{\mu\nu}\partial_\mu F,\partial_\nu F=0.
\]
For the class of solutions considered here, this condition reduces to
\begin{eqnarray}
  r^2 r_H'(\theta)^2+{ H}(r,\theta)\vert_{r=r_H(\theta)}=0,\qquad r_H'(0)=r_H'(\pi)=0,
  \label{num}
\end{eqnarray}
where ${H}(r,\theta)=(r^2+a^2)(r^2+b^2)-Mr^2-r^2\alpha \Sigma c_{4,1}h_2(r,\theta)^2$. Nevertheless, before implementing the numerical analysis, one must first ensure the existence of a physically acceptable horizon. In particular, the function ${\cal H}(r,\theta)$ appearing in the metric ansatz must possess at least one root, since otherwise no regular horizon structure can arise and consequently no black-hole solution exists within the present setup. We therefore start by studying the conditions under which ${\cal H}(r,\theta)$ admits a root and determine the restrictions imposed by this requirement.

Once the existence of a suitable horizon configuration is established, the numerical construction proceeds according to the same strategy in all cases considered here. The first step consists in locating the turning points by solving simultaneously
\[
{H}(r,\theta)=0,
\qquad
r_H'(\theta)=0,
\]
which determine the critical points of the system and provide the relevant initial configurations. Around these points, a local quadratic expansion is constructed in the form
\[
r_H(\theta)\simeq r_0+h_2\theta^2,
\]
where $ r_0$ is a constant and, where the branch associated with the root of smallest absolute value is selected as the appropriate initial seed \footnote{We stress that this selection concerns the two roots of the \emph{local expansion coefficient} $h_2$, not a choice between candidate values of $r_0$ itself: the turning point $r_0$ is always fixed beforehand as the unique outermost root of $H(r,\theta)=0$, as established above via $r_+(\theta)=\sup\mathcal{Z}_\theta$. Substituting the quadratic ansatz into Eq.~\eqref{num} and expanding to leading order around the turning point $(r_0,\theta_0)$ yields, after dividing through by $r_0^2$, a quadratic equation for $h_2$ of the schematic form $4h_2^2+\partial_r\!\left(H/r^2\right)h_2+\tfrac12\partial_\theta^2\!\left(H/r^2\right)=0$, evaluated at $(r_0,\theta_0)$, which admits two real roots in the cases considered. Only the root of smaller magnitude reproduces, upon integration of the equivalent second-order equation, a solution that matches the independently computed value of $r_H$ at the neighbouring turning point; the other root corresponds to an unstable branch of the same local expansion that fails to connect smoothly to the rest of the domain. The selection therefore has no bearing on which horizon (inner or outer) is being described, but rather on which of the two locally admissible continuations of the (already outermost) horizon curve is globally consistent.}. Finally, the equivalent second-order differential equation is integrated from the stable endpoint, allowing the complete numerical profile of the solution to be obtained.

As a first step, let us ensure that the equation ${H}(r,\theta)=0$ always admits at least one positive root. We begin by considering the reduced solution obtained from \eqref{KSansatz22} by setting $c_{31}=c_{42}=0$. This leads to a quartic equation for the radial coordinate, whose $\theta$-dependent root of 
\begin{equation}
    (r^2+a^2)(r^2+b^2) - r^2 \left( M + \alpha c_{41} \mathcal{F}_1(\theta)^2 \right) = 0,
\end{equation}
is explicitly given by
\begin{equation}
    \frac{1}{\sqrt{2}} \sqrt{ M + \alpha c_{41} \mathcal{F}_1(\theta)^2 - (a^2+b^2) + \sqrt{ \left[ M + \alpha c_{41} \mathcal{F}_1(\theta)^2 - (a+b)^2 \right] \left[ M + \alpha c_{41} \mathcal{F}_1(\theta)^2 - (a-b)^2 \right] } }.
\end{equation}
The reality condition of this root imposes the constraint
\begin{equation}
    M + \alpha c_{41} \mathcal{F}_1(\theta)^2 \geq \max\Big\{(a+b)^2,\,(a-b)^2\Big\}, \qquad \forall\,\theta \in [0,\pi].
\end{equation}
This condition ensures the positivity of the argument of the square root, thereby excluding the formation of naked singularities. Owing to the functional freedom in the choice of $\mathcal{F}_1(\theta)$, this inequality can always be satisfied by an appropriate tuning of the Proca field profile.

A similar analysis can be carried out for the solution \eqref{dfg} with 
\[
H(r,\theta):= (r^2+a^2)(r^2+b^2) - M r^2 - r^2 \alpha c_{41} \mathcal{F}_1(\theta)^2
\left(\frac{\Sigma}{\Sigma+2r^2}\right)^{\frac{2c_{42}}{2c_{42}+c_{41}^2}} = 0.
\]
The asymptotic behaviour of the function is readily obtained as
\[
H(r,\theta)\underset{r\to\infty}{\sim} r^4 > 0,
\]
which ensures that $H(r,\theta)$ is positive for sufficiently large values of $r$. Moreover, by choosing for instance
\begin{equation}
    \mathcal{F}_1(\theta)=\sqrt{\frac{(1+a^2)(1+b^2)}{\alpha c_{41}}}
\left[\frac{(a^2-b^2)\cos^2\theta+3+b^2}{(a^2-b^2)\cos^2\theta+1+b^2}\right]^{\frac{c_{42}}{2c_{42}+c_{41}^2}},\label{F1choice}
\end{equation}
one obtains (provided that $M>0$)
\[
H(1,\theta) = -M < 0, \qquad \forall\,\theta.
\]
Since $H(r,\theta)$ is continuous with respect to $r$, the intermediate value theorem ensures that, for each fixed $\theta$, the equation $H(r,\theta)=0$ admits at least one solution in $(1,\infty)$. To further characterize the set of roots, we analyse the radial derivative. One finds that
\[
\partial_r H(r,\theta) \xrightarrow[r\to\infty]{} +\infty,
\]
independently of $\theta$. Consequently, there exists $R>0$ such that
\[
\partial_r H(r,\theta) > 0, \qquad \forall r \ge R, \quad \forall \theta,
\]
so that $H(r,\theta)$ is strictly increasing on $(R,\infty)$. Let
\[
\mathcal{Z}_\theta = \{ r>1 \mid H(r,\theta)=0 \}.
\]
From the previous discussion, $\mathcal{Z}_\theta$ is non-empty. Furthermore, since $H(r,\theta)$ is strictly increasing for $r \ge R$, it admits at most one zero in $(R,\infty)$. It follows that $\mathcal{Z}_\theta$ is bounded from above, and we can define
\[
r_+(\theta) := \sup \mathcal{Z}_\theta.
\]
By continuity of $H$, one has $H(r_+(\theta),\theta)=0$, so that $r_+(\theta)$ is the largest root of the equation. Finally, due to the strict monotonicity of $H$ on $(R,\infty)$, it admits at most one zero in this interval. In particular, if a root exists in $(R,\infty)$, it is unique. Otherwise, all the roots lie in the interval $(1,R]$, and $r_+(\theta)$ remains well-defined as the largest root, although no uniqueness can be inferred in this region. (See Fig. 1) \\

\begin{figure}[h!]
\centering
\includegraphics[width=0.48\textwidth]{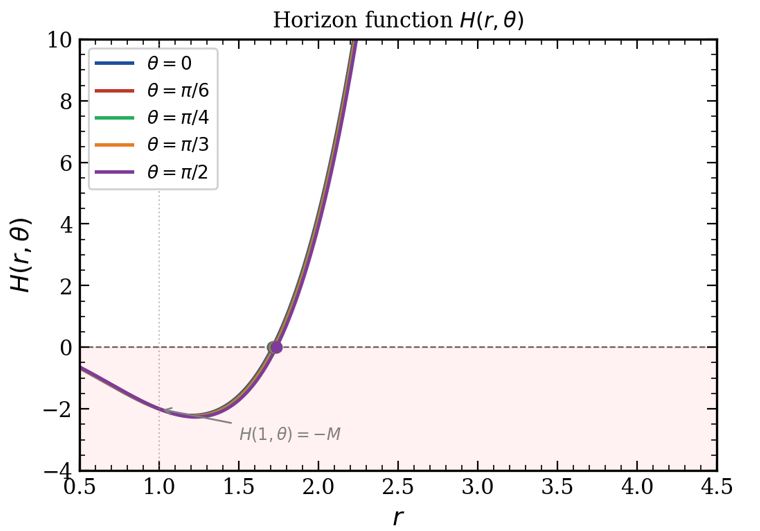}
\hfill
\includegraphics[width=0.48\textwidth]{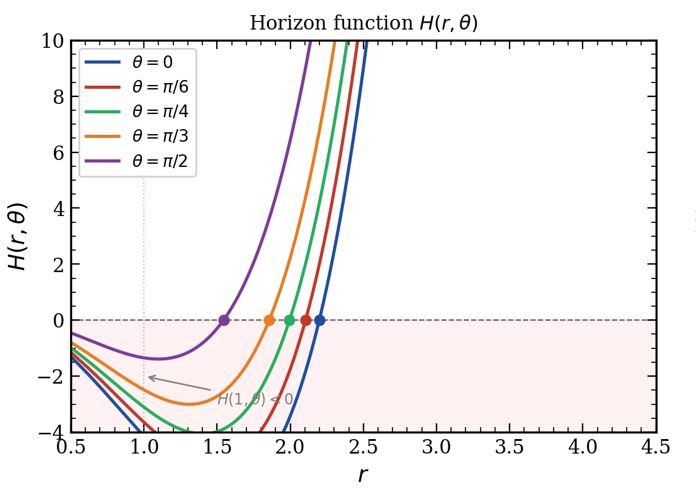}
\caption{Root of the function $H(r,\theta)$ for five values of the polar 
angle $\theta \in \{0,\pi/6,\pi/4,\pi/3,\pi/2\}$, with parameters 
$a=0.8$, $b=0.5$, $M=2$, $\alpha=c_{41}=1$, $c_{42}=0.5$. 
\textit{Left panel:} Analytic choice of $\mathcal{F}_1(\theta)$ 
from Eq.~\eqref{F1choice}, for which $H(1,\theta)=-M$ exactly for 
all $\theta$ (vertical dotted line at $r=1$), demonstrating that 
the hypotheses of the intermediate value theorem are satisfied. 
\textit{Right panel:} Illustrative choice $\mathcal{F}_1(\theta) = 
1.5 + \cos\theta$, showing a pronounced angular dependence of the 
outer horizon $r_+(\theta)$ (filled circles). 
In both cases, the shaded region indicates $H < 0$.}
\label{fig:horizon}
\end{figure}

\begin{figure}[h!]
\centering
\includegraphics[width=0.48\textwidth]{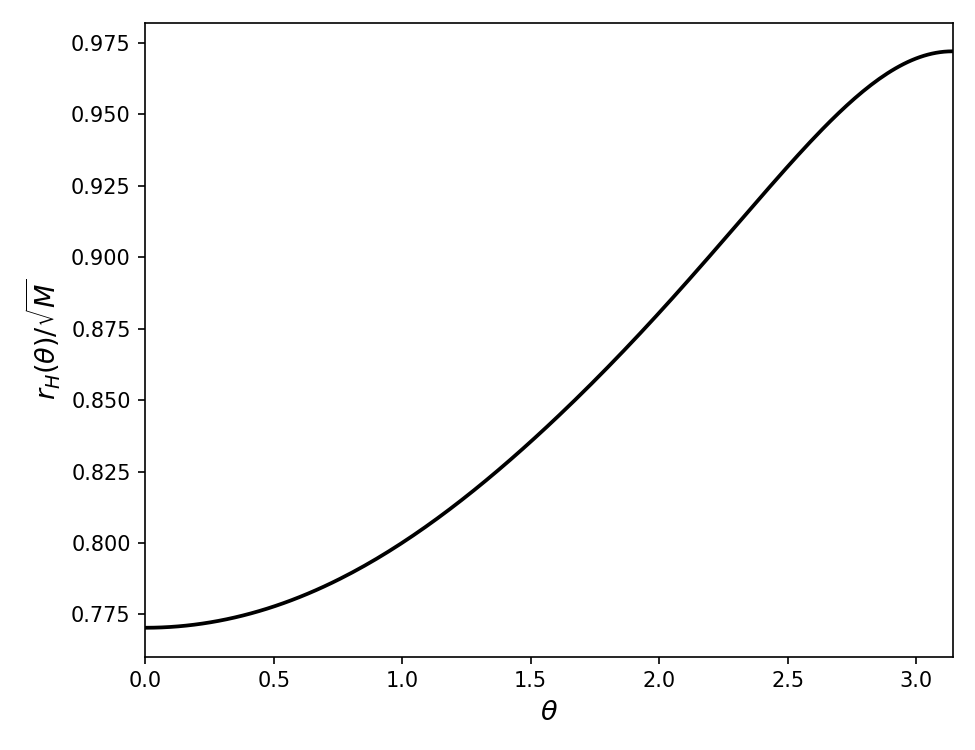}
\hfill
\includegraphics[width=0.48\textwidth]{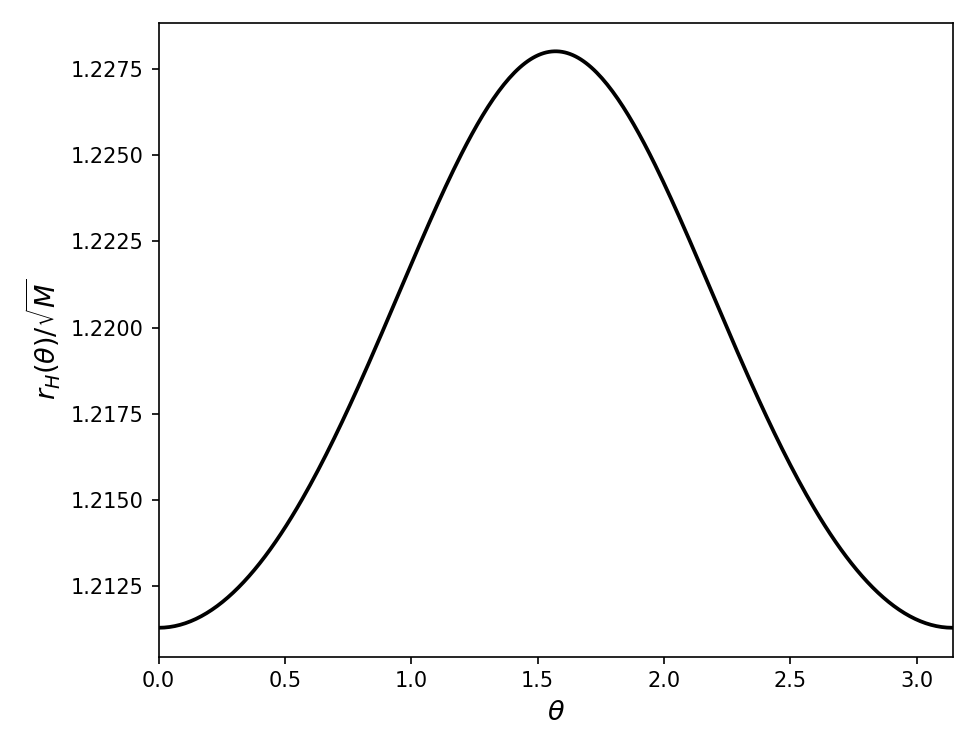}
\caption{Non-circular horizon location $r_H(\theta)/\sqrt{M}$ as a function of the polar angle $\theta\in[0,\pi]$, obtained by numerically solving Eq.~\eqref{num}, for two of the solution branches discussed in the text, with parameters $a=0.8$, $b=0.5$, $M=2$, $\alpha=c_{4,1}=1$. \textit{Left panel:} solution branch of Eq.~\eqref{h2casoI} (Case I), with $c_{3,1}=-4$ and the illustrative, non-symmetric choice $\mathcal{F}_1(\theta)=1+0.5\cos\theta$. Since $\mathcal{F}_1(\theta)\neq\mathcal{F}_1(\pi-\theta)$, the equatorial symmetry is broken and the profile is monotonic, with $r_H(0)=1.089434$ and $r_H(\pi)=1.374741$, and no interior turning point. \textit{Right panel:} solution branch of Eq.~\eqref{dfg} (Case III), with $c_{4,2}=0.5$ and the analytic choice of $\mathcal{F}_1(\theta)$ from Eq.~\eqref{F1choice}, for which $H(1,\theta)=-M$ exactly for all $\theta$. This choice satisfies $\mathcal{F}_1(\theta)=\mathcal{F}_1(\pi-\theta)$, preserving the equatorial symmetry and yielding $r_H(0)=r_H(\pi)=1.713028$, with a maximum at the equator $r_H(\pi/2)=1.736674$. In both panels, the profile was obtained by numerically integrating the second-order differential equation equivalent to Eq.~\eqref{num}, starting from the stable endpoint using the quadratic, minimal-magnitude local expansion described in the text.}
\label{fig:horizonIyIII}
\end{figure}
\clearpage

Having established the existence of a root for the function $H(r,\theta)$, we now proceed to analyse the horizon equation numerically. To this end, we have generated several plots in order to verify that the equation \eqref{num} defining the horizon admits consistent numerical solutions within the parameter ranges considered, (see Fig. 2).

\subsubsection{Toroidal case}

Another noteworthy observation is that these solutions admit a natural extension to the toroidal case ($\kappa=0$) within the Kerr--Schild framework. The systematic study of toroidal, and more generally topological, black holes was originally developed in four-dimensional Anti-de Sitter (AdS) gravity \cite{Lemos:1994xp, Klemm:1997ea, Birmingham:1998nr}. These works demonstrated that the sign of the sectional curvature $\kappa$ plays a fundamental role in shaping both the causal structure and the thermodynamic properties of the resulting spacetimes. Extending this perspective to five dimensions provides a non-trivial arena to test the robustness of the primary Proca hair under modifications of the horizon topology and asymptotic structure. In the planar case $\kappa=0$, the seed metric $ds_0^2$ corresponds to a flat Lorentzian manifold endowed with toroidal topology, and the null geodesic vector $\ell$ is naturally adapted to Cartesian-like coordinates on the torus. The Kerr--Schild ansatz then takes the form
\begin{eqnarray}
&&ds^2 = -(-\lambda r^2)dt^2 + \frac{dr^2}{(-\lambda r^2)} + r^2\left(d\theta^2 + dx^2 + dy^2\right) + h_1(r,\theta)\,\ell \otimes \ell, \qquad A = h_2(r,\theta)\,\ell, \nonumber\\
&&\ell = \sqrt{1-\mathrm{sgn}(\lambda)(a^2+b^2)}\,dt + \frac{dr}{(-\lambda r^2)} - \frac{1}{\sqrt{|\lambda|}}\left(a\,dx + b\,dy\right).
\end{eqnarray}
Substituting this ansatz into the field equations, one obtains
\begin{equation}
    h_1(r,\theta) = \frac{M}{r^2} + \frac{h_2(r,\theta)^2\,G_{4X}}{G_{4}}, \label{h_1eqk0}
\end{equation}
so that the metric can be recast in the compact Kerr--Schild form
\begin{equation}
\label{KSansatz2k0}
d s^2 = d s_0^2 + \frac{M}{r^2}\, \ell \otimes \ell + \frac{G_{4X}}{G_4} A \otimes A, 
\qquad A = h_2(r, \theta)\, \ell.
\end{equation}
Within this geometry, the remaining non-trivial field equation, namely $\mathcal{J}^{r}=0$, remarkably reduces to a total derivative,
\begin{equation}
    \mathcal{J}^{r}= \frac{\partial}{\partial r} \left\{ 
    \frac{3 r^2 \left( G_{4X}^2 + G_{4XX} G_{4} \right) h_2^2 
    + G_{3X} h_2 r^3 G_{4} 
    + 3 G_{4X} {M} G_{4}}{G_{4}} 
    \right\}. \label{jk0eval} 
\end{equation}
This structure allows for a direct integration of the field equations. Making use of the perturbative expansions of the coupling functions $G_n(X)$ introduced in \eqref{Gn_expansions_final}, one can determine the radial profile $h_2(r,\theta)$ explicitly. The general solution is found to be
\begin{equation}
    h_2(r,\theta) = \frac{\sqrt{r^4 c_{31}^2 \mathcal{F}_1^2(\theta) + 12 \alpha (2 c_{42} + c_{41}^2)} - r^2 c_{31} \mathcal{F}_1(\theta)}{6 \alpha (2 c_{42} + c_{41}^2)\, \mathcal{F}_1(\theta)\, r}.
\end{equation}
The validity of this expression requires the denominator to be non-vanishing, which imposes the condition $c_{42} \neq -c_{41}^2/2$. This special parameter choice corresponds to a bifurcation point in the solution space and must therefore be treated separately. Indeed, setting $c_{42} = -c_{41}^2/2$ and analysing the field equations, one finds that the profile function simplifies considerably, yielding a power-law behaviour of the form
\begin{equation}
    h_2(r,\theta) = \frac{\mathcal{F}_1(\theta)}{r^3}.
\end{equation}
In both branches of the solution space, the function $\mathcal{F}_1(\theta)$ arises as an arbitrary integration function depending solely on the angular coordinate, reflecting the persistence of a non-trivial angular structure associated with the Proca hair.

\subsubsection{Maxwell obstruction }

To end this section, we look to the action \eqref{procaLgeneral} supplemented by the Maxwell term  \( \epsilon\,F_{\mu\nu}F^{\mu\nu} \) by adopting the same ansatz \eqref{KSansatz}. Without any loss of generality we can restrict ourselves to the case without cosmological constant and with $G_2=0$. In this case, the equations of motion for the gauge field \( A_\mu \) are modified accordingly. In particular, the radial component of the current becomes
\begin{align*}
\mathcal{J}^{r}_\epsilon 
= \mathcal{J}^{r} 
+ \epsilon\,\frac{\partial}{\partial r}\Big(P(r,\theta)\Big)\,\nabla_{\nu}(F^{r\nu}) \, ,
\end{align*}
where $ \mathcal{J}^{r}$ and $P(r,\theta)$ are  defined respectively in \eqref{Jrr} and in \eqref{prt}. On the other hand, new components of the vector field equation $\mathcal{J}^{\mu}=0$ arise, and are given by 
\begin{align*}
\mathcal{J}^{\phi}_\epsilon+\frac{a}{r^2+a^2}\mathcal{J}^{r}_\epsilon
&\propto \epsilon\left(\nabla_{\nu}(F^{\phi\nu})+\frac{a}{r^2+a^2}\nabla_{\nu}(F^{r\nu})\right),\\
\mathcal{J}^{\psi}_\epsilon+\frac{b}{r^2+b^2}\mathcal{J}^{r}_\epsilon
&\propto \epsilon\left(\nabla_{\nu}(F^{\psi\nu})+\frac{b}{r^2+b^2}\nabla_{\nu}(F^{r\nu})\right),\\
\mathcal{J}^{t}_\epsilon+\mathcal{J}^{r}_\epsilon
&\propto \epsilon\,\nabla_{\nu}(F^{t\nu}+F^{r\nu}),\qquad \mathcal{J}^{\theta}_\epsilon\propto \epsilon\,\nabla_{\nu}(F^{\theta\nu}) \, .
\end{align*}
From these equations, and considering suitable linear combinations among them, it is possible to isolate a relation that determines the function \( h_2(r,\theta) \) valid for $\epsilon\not=0$ as
\begin{equation*}
h_2(r,\theta) = -\frac{Q}{\Sigma} \, .
\end{equation*}
As well, the equations arising from the Einstein sector are also modified. In particular, the relevant equations take the following form
\begin{align*}
E_{rr_\epsilon} 
&:= \frac{\partial}{\partial r}\left[
\Sigma\,\left(h_2^2\,G_{4X}-G_{4}\,h_1
\right)\right]
+\frac{\epsilon r}{2(\Sigma+2r^2)}
\left(
\Sigma^2\frac{\partial}{\partial r}\big(h_2\big)^2
-4\,h_2^2\,(r^2-\Sigma)
\right),\\[6pt]
E_{\theta\theta_\epsilon} 
&:= \frac{\partial}{\partial \theta}\left[
\Sigma\,
\Bigl(
h_2^2\,G_{4X}-G_{4}\,h_1
\Bigr)
\right]
-\epsilon\,h_2\,\frac{\partial}{\partial \theta}
\left(\Sigma\,h_2\right) \, .
\end{align*}
It is important to observe that, upon substituting the previously obtained expression for \( h_2 \) into these last two equations, the compatibility condition $\partial_{\theta}E_{rr_\epsilon}-\partial_rE_{\theta\theta_\epsilon} =0 $ is valid only if $Q=0$. Therefore, the standard Maxwell framework is strictly incompatible with this setup, as consistency requires turning off the function $h_2$, which equates to a trivial vector field. In the static case, it was shown in \cite{Cisterna:2016nwq} that this restriction can be circumvented by considering non-linear electrodynamics of the form $\left(-F_{\mu\nu}F^{\mu\nu}\right)^p$, \cite{Hassaine:2007py, Hassaine:2008pw}. Motivated by this, we investigate whether introducing such a non-linear Maxwell framework can similarly restore compatibility when rotation is included. However, as we demonstrate below, this mechanism fails to produce consistent non-trivial solutions in the rotating regime. So let us consider the following Lagrangian 
\[
{\cal L}=\epsilon\,\, \left(-F_{\mu\nu}F^{\mu\nu}\right)^p,\qquad p\not=1.
\]
As in the standard Maxwell case, the non-trivial $\mathcal{J}^{\mu}=0$ equations are given by 
\begin{align*}
\mathcal{J}^{r}_\epsilon 
&= \mathcal{J}^{r} 
+ \epsilon\,\frac{\partial}{\partial r}\Big(P(r,\theta)\Big)\,\nabla_{\nu}(F^{r\nu}\mathcal{F}^p) \, ,
\\
\mathcal{J}^{\phi}_\epsilon+\frac{a}{r^2+a^2}\mathcal{J}^{r}_\epsilon
&\propto \epsilon\left(\nabla_{\nu}(F^{\phi\nu}\mathcal{F}^p)+\frac{a}{r^2+a^2}\nabla_{\nu}(F^{r\nu}\mathcal{F}^p)\right),\\
\mathcal{J}^{\psi}_\epsilon+\frac{b}{r^2+b^2}\mathcal{J}^{r}_\epsilon
&\propto \epsilon\left(\nabla_{\nu}(F^{\psi\nu}\mathcal{F}^p)+\frac{b}{r^2+b^2}\nabla_{\nu}(F^{r\nu}\,\mathcal{F}^p)\right),\\
\mathcal{J}^{t}_\epsilon+\mathcal{J}^{r}_\epsilon
&\propto \epsilon\,\nabla_{\nu}([F^{t\nu}+F^{r\nu}]\mathcal{F}^p),\qquad \mathcal{J}^{\theta}_\epsilon\propto \epsilon\,\nabla_{\nu}(F^{\theta\nu}\mathcal{F}^p) \, .
\end{align*}
where $\mathcal{F}=F_{\mu\nu}F^{\mu \nu}$. Now, as  in the previous case, an appropriate linear combination of the generalized Maxwell equation yields the following equation
\begin{equation} 
\left[\Sigma^2\left(\partial_r h_2\right)^2 
- 4\left(\Sigma - r^2\right)h_2^2\right]^{p-1} 
\, \left(\Sigma \, \partial_r h_2 + 2r \, h_2\right)=0.
\end{equation}
This equation admits two non-trivial solutions 
which arise from setting each factor to zero independently. The first is 
obtained by requiring $\Sigma\,\partial_r h_2 + 2r\,h_2 = 0$, which 
recovers the solution found previously, this is 
$h_2 = \mathcal{F}_1(\theta)/\Sigma$, where the remaining Maxwell equations 
further constrain $\mathcal{F}_1(\theta)$ to be a constant. The second 
non-trivial branch $p\not=1$ arises from the vanishing of the first factor, 
yielding
\begin{equation}
    h_2(r,\theta) = \mathcal{F}_1(\theta) \,e^{\pm2\arctan\left(\frac{r}{\sqrt{\Sigma(r,\theta) - r^2}}\right)}. \label{rama3}
\end{equation}
On the other hand, the master equations take the form
\begin{equation}
\begin{aligned}
E_{\theta\theta_\epsilon} :=& \frac{\partial}{\partial \theta}\left[
\Sigma\,
\Bigl(
h_2^2\,G_{4X}-G_{4}\,h_1
\Bigr)
\right]
+\epsilon\,(-1)^p \,p  2^{(p-1)}\,h_2\, 
     \,\left[\Sigma^2\left(\partial_r h_2\right)^2 
- 4\left(\Sigma-r^2\right)h_2^2\right]^{p-1}\,\frac{\partial}{\partial \theta}
\left(\Sigma\,h_2\right) \,,\\[6pt]
   E_{rr_\epsilon} 
:=& \frac{\partial}{\partial r}\left[
\Sigma\,\left(h_2^2\,G_{4X}-G_{4}\,h_1
\right)\right]+\\
&\epsilon \frac{(-1)^{p+1} \, (2p-1)  2^{p-2} \, r 
\, \left[\Sigma^2\left(\frac{\partial h_2}{\partial r}\right)^2 
- 4\left(\Sigma - r^2\right)h_2^2\right]^{p-1} 
\, \left[\Sigma^2\left(\frac{\partial h_2}{\partial r}\right)^2 
- \dfrac{4}{2p-1}\left(\Sigma - r^2\right)h_2^2\right]}
{(\Sigma + 2r^2)  \Sigma^{2(p-1)}}, 
\end{aligned}
\end{equation}
It is worth noting that the $\epsilon$-dependent terms in both 
equations contain precisely the factor that vanishes for the solution 
found above, see Eq.~\eqref{rama3}. Consequently, these terms do 
not contribute, and the equation for $h_1$ reduces to (\ref{h_1eq}). On the other hand, since
\begin{equation}
\mathcal{J}^{r}_\epsilon 
= \mathcal{J}^{r} 
+ \epsilon\,\frac{\partial}{\partial r}\Big(P(r,\theta)\Big)\,
\nabla_{\nu}(F^{r\nu}\mathcal{F}^p),
\end{equation}
the expression $h_2 =\mathcal{F}_1(\theta)\,e^{\pm 2\arctan\!\left(r/\sqrt{\Sigma-r^2}\right)}$ 
cancels the $\epsilon$-dependent contribution identically, as shown above. 
However, the remaining piece $\mathcal{J}^{r}=0$ must still be satisfied 
independently. Evaluating this equation with 
\begin{equation*}
h_1(r,\theta) = \frac{M}{\Sigma} + \frac{h_2(r,\theta)^2\,G_{4X}}{G_{4}},
\end{equation*}
one finds that none of the analytic solutions obtained in 
Section~\ref{subsec:Analytic} for $\mathcal{J}^{r}=0$ are compatible 
with the exponential form of $h_2$ above \eqref{rama3}. Consequently, this branch 
does not yield a consistent solution within the generalized Proca 
theory for generic coupling functions, and must therefore be discarded.

\section{Conclusion and Perspectives}

In this work, we have constructed a new class of exact rotating black hole solutions in five-dimensional generalized Proca theories. The key ingredient is a Kerr--Schild ansatz in which the Proca vector field is aligned with a null geodesic congruence as done in \cite{Fernandes:2026rjs}. This choice considerably simplifies the field equations, reducing them to a set of three coupled equations that can be solved analytically. In particular, the alignment ensures that the vector field is light-like on-shell, so that the functional couplings in the Proca Lagrangian effectively behave as constants, allowing us to bypass much of the usual technical difficulty.

The solutions describe black holes with two independent angular momenta and can be constructed for different horizon topologies, including the spherical ($\kappa=1$) and planar ($\kappa=0$) cases. Obtaining analytic rotating solutions in higher dimensions with multiple rotation parameters is typically very challenging, especially in modified gravity where the field equations are highly non-linear and often lead to coupled partial differential equations. In many situations, one has to rely on numerical methods or perturbative approaches. In contrast, the Kerr--Schild framework used here allows for a fully analytic treatment, even in the presence of a cosmological constant, illustrating its effectiveness for exploring non-linear gravitational configurations.

A distinctive and particularly intriguing feature of these solutions is the emergence of primary hair, in close analogy with what has been observed in four-dimensional Proca models \cite{Fernandes:2026rjs}. All the solution branches identified in this work, with the exception of the stealth configuration, are characterized by the presence of an arbitrary integration function $\mathcal{F}_1(\theta)$. This functional freedom arises naturally from the structure of the reduced field equations, which are effectively radial and therefore do not fully constrain the angular dependence. Beyond its mathematical origin, this arbitrariness plays an important role at the physical level. In particular, it provides a mechanism to ensure the existence of a regular event horizon. As shown in the previous analysis, by an appropriate choice of the angular profile $\mathcal{F}_1(\theta)$, one can guarantee the existence of an event horizon radius. In this sense, the Proca hair is not merely a by-product of the integration procedure, but rather a crucial ingredient in the construction of physically acceptable black hole geometries. At the same time, the presence of this arbitrary function points to a non-trivial degeneracy in the space of solutions, which is absent in standard General Relativity. Different choices of $\mathcal{F}_1(\theta)$ may lead to geometries with distinct physical properties, raising the question of whether all such configurations should be regarded as physically equivalent. Clarifying the status of this "hair" therefore requires further investigation. In particular, it will be important to determine whether $\mathcal{F}_1(\theta)$ encodes a genuine physical degree of freedom, or whether it can be constrained, or even completely fixed, by imposing additional physical requirements, such as regularity at the horizon, stability under linear perturbations, or suitable asymptotic fall-off conditions. Understanding these aspects is essential for assessing the physical relevance of the solutions presented here and for establishing whether the associated Proca hair can play an observable role in higher-dimensional gravitational systems.

Several natural directions for future work follow from our results. A first step would be to extend these solutions to an arbitrary number of dimensions, in order to make contact with the general Myers-Perry family and obtain a broader picture of rotating Proca black holes in higher dimensions. In addition, since our solutions are five-dimensional, it would be interesting to explore them in the context of the $\text{AdS}_5/\text{CFT}_4$ correspondence \cite{Maldacena:1997re}. In particular, one could investigate how these configurations with Proca hair are encoded in the dual field theory and whether they lead to new features on the holographic side. Another natural direction for future work arises from the non-circular nature of our rotating Proca solutions. It would be of great interest to investigate whether the Proca vector field $\mathcal{A}_\mu$ could be employed to generate a disformal transformation of the metric, $\tilde{g}_{\mu\nu} = g_{\mu\nu} + \phi\mathcal{A}_\mu \mathcal{A}_\nu$, in order to bring the spacetime into a circular form. This approach is motivated by the scalar field case; while starting from a stealth configuration on a Kerr background \cite{Charmousis:2019vnf} generally breaks circularity \cite{Anson:2020trg, BenAchour:2020fgy}, there exist other scalar field solutions where the circularity of the metric can be preserved or restored through suitable disformal mappings \cite{BenAchour:2025lkx}. Applying a similar logic to the Proca case would allow us to determine if the vector field can be tuned to restore circularity, potentially leading to new classes of exact rotating solutions. Also, our analysis has shown that the inclusion of a standard Maxwell term in these generalized Proca theories, whether in four or five dimensions, appears to be incompatible with the Kerr-Schild ansatz. While we extended this investigation to include non-linear electrodynamics of the $(F_{\mu\nu}F^{\mu\nu})^p$ type, the results remained similarly constrained. Consequently, a promising next step would be to explore a broader class of frameworks, such as the recently classified Degenerate Higher-Order Maxwell-Einstein (DHOME) theories \cite{Colleaux:2023cqu, Colleaux:2024ndy, Colleaux:2025vtm}. It remains to be seen whether specific sub-sectors within this classification might accommodate such a construction while preserving the consistency of the Kerr-Schild framework.

Finally, the success of the Kerr-Schild ansatz in the generalized Proca sector provides a powerful proof of concept for other alternative theories of gravity. In particular, scalar-tensor theories, where the search for exact rotating solutions has proven to be exceedingly tricky and often elusive, could benefit from a similar approach. Understanding the precise limits and conditions under which the Kerr-Schild ansatz "works" in modified gravity will be essential for discovering new analytic backgrounds, which are indispensable for testing the limits of modified gravity theories.

\section*{Acknowledgments}
The work of M.~H. is partially supported by
FONDECYT grant 1260479. U. H. V. thanks CINVESTAV, Mexico, for its warm hospitality  during the development of this work.


\bibliography{References}

\end{document}